\documentclass[10pt,aps,prl,twocolumn,groupedaddress,showkeys]{revtex4-2}
\usepackage{amsmath}
\usepackage[english]{babel}
\usepackage{graphicx}

\DeclareGraphicsExtensions{.pdf,.eps,.png,.jpg,.mps}
\usepackage{epstopdf}
\usepackage{threeparttable}
\usepackage{color}
\usepackage[T2A]{fontenc}
\usepackage[utf8]{inputenc}
\usepackage{float}
\usepackage{tabularx}
\usepackage{array}
\usepackage[font=small,labelfont=bf]{caption}
\usepackage[percent]{overpic}
\usepackage{multirow}
\usepackage[colorlinks=true,linkcolor=blue,urlcolor=blue,citecolor=blue,anchorcolor=blue]{hyperref}

\begin{document}

\title{Highly Efficient Exciton Modulation in MoSe$_2$/PdSe$_2$ Heterostructures}
\author{Petr Rozhin$^1$}
\author{Emma Contin$^1$}
\author{Danae Katrisioti$^{2,3}$}
 \author{Till Weickhardt$^4$}
\author{Muhammad Sufyan Ramzan$^5$}
\author{Micol Bertolotti$^6$}
\author{Nouha Loudhaief$^1$}
\author{Bing Wu$^7$}
\author{Zdeněk Sofer$^7$}
\author{Takashi Taniguchi$^8$}
\author{Kenji Watanabe$^9$}
\author{Leonardo Puppulin$^1$}
\author{Stefano Dal Conte$^6$}
\author{Caterina Cocchi$^{5}$}
\author{Ioannis Paradisanos$^{2,3}$}
\author{Giancarlo Soavi$^4$}
\author{Giovanni Antonio Salvatore$^1$}
\author{Domenico De Fazio$^1$}
\email[]{domenico.defazio@unive.it}
\affiliation{$^1$ Department of Molecular Sciences and Nanosystems, Ca’ Foscari University of Venice, 30172 Venice, Italy}
\affiliation{$^2$ Department of Materials Science and Engineering, University of Crete, Heraklion, 70013 Crete, Greece}
\affiliation{$^3$ Institute of Electronic Structure and Laser, Foundation for Research and Technology-Hellas, Heraklion, 71110 Crete, Greece}
\affiliation{$^4$ Institute of Solid State Physics, Friedrich Schiller University Jena, Jena 07743, Germany}
\affiliation{$^5$ Institute of Condensed-Matter Physics and Optics, Friedrich Schiller University Jena, Jena 07743, Germany}
\affiliation{$^6$ Department of Physics, Politecnico di Milano, I-20133 Milano, Italy}
\affiliation{$^7$ Department of Inorganic Chemistry, University of Chemistry and Technology Prague, 166 28 Prague 6, Czech Republic}
\affiliation{$^8$ Research Center for Materials Nanoarchitectonics, National Institute for Materials Science, 1-1 Namiki, Tsukuba 305-0044, Japan}
\affiliation{$^9$ Research Center for Electronic and Optical Materials, National Institute for Materials Science, 1-1 Namiki, Tsukuba 305-0044, Japan}


\begin{abstract}
Controlling exciton recombination in atomically thin semiconductors is central to their optoelectronic functionality, as the competition between radiative and non-radiative decay channels governs emission efficiency. Existing approaches, such as defect passivation, chemical doping, dielectric engineering, and strain tuning, primarily aim to suppress non-radiative losses. Here, we report a pronounced $\sim$6-fold enhancement of room-temperature A-exciton emission in a type-I MoSe$_2$/PdSe$_2$ van der Waals heterostructure, yielding a photoluminescence quantum yield of 6\%, compared to $\sim$1\% for as-exfoliated monolayer MoSe$_2$. This enhancement is accompanied by strong quenching of the B-exciton, consistent with interlayer electronic coupling that redistributes exciton populations toward the radiative A-exciton channel. Power- and temperature-dependent measurements reveal a suppression of exciton–exciton annihilation and a crossover to quenched emission at low temperature, indicating a redistribution of exciton relaxation pathways. Photoluminescence excitation spectroscopy further reveals a broadband enhancement spanning 450–725 nm, ruling out a resonance-specific mechanism. These results demonstrate that interlayer electronic coupling can be used as an efficient means to redirect exciton populations toward radiative channels, enhancing emission efficiency in two-dimensional semiconductors without chemical modification or strain.
\end{abstract}

\maketitle

\section{\label{Intro}Introduction}
Strong, stable light emission is essential for optoelectronic devices such as light-emitting diodes (LEDs)~\cite{wang_highly-efficient_2020}, lasers~\cite{salehzadeh_optically_2015}, and single-photon sources~\cite{aharonovich_solid-state_2016}. Efficient emission occurs when excitons recombine radiatively, consistently producing photons, rather than being lost to heat~\cite{kim_inhibited_2021}, defect trapping~\cite{reshchikov_mechanisms_2021}, phonon-assisted decay~\cite{valerini_temperature_2005}, and Auger recombination~\cite{shen_auger_2007, zheng_light_2018}. Suppression of these non-radiative pathways increases the photoluminescence quantum yield (PLQY), defined as the ratio of emitted photons to absorbed photons, thereby improving the efficiency of light-emitting devices and reducing their power consumption, both key aspects in modern photonic components~\cite{ross_electrically_2014, lin_perovskite_2018, bai_perovskite_2023, tuong_ly_near-infrared_2017}.

Transition metal dichalcogenides (TMDs) such as MoS$_2$, MoSe$_2$, WS$_2$, and WSe$_2$, have emerged as promising platforms for next-generation optoelectronics owing to their unique properties, including a direct bandgap in the monolayer limit~\cite{zhao_electronic_2015}, strong photoluminescence (PL)~\cite{splendiani_emerging_2010}, large exciton binding energies (typically $\sim$300–600 meV)~\cite{kylanpaa_binding_2015, mueller_exciton_2018}, pronounced spin-orbit coupling~\cite{xiao_coupled_2012}, mechanical flexibility~\cite{gao_flexible_2017}, and the ability to form van der Waals (vdW) heterostructures with novel functionalities~\cite{li_devices_2017, geim_van_2013}. Furthermore, their atomic thickness makes their optical response highly sensitive to the surrounding dielectric environment~\cite{raja_coulomb_2017, chernikov_exciton_2014}. While this sensitivity enables powerful control of excitonic properties through dielectric engineering, it also makes emission strongly dependent on defects~\cite{amani_near-unity_2015}, substrate disorder~\cite{rhodes_disorder_2019}, and interface quality~\cite{cadiz_excitonic_2017}. 

Several approaches have been employed to enhance the PLQY of monolayer TMDs, including encapsulation~\cite{cadiz_excitonic_2017, wierzbowski_direct_2017, ryu_optical_2024, li_macroscopic_2023}, chemical passivation~\cite{amani_near-unity_2015, tanoh_giant_2021, li_mechanistic_2021}, and strain engineering~\cite{li_optoelectronic_2015, lee_drift-dominant_2022}. Encapsulation with dodecanol passivates Se vacancies and suppresses substrate-induced non-radiative recombination, enabling PLQY values from 6.2\% up to $\sim$22\% at 5 K in MoSe$_2$~\cite{li_macroscopic_2023}. In addition, the thickness of the hexagonal boron nitride (hBN) encapsulation layer can control the exciton radiative rate via the Purcell effect (where the local optical environment alters the spontaneous emission rate of excitons) ~\cite{PhysRevLett.123.067401}. Chemical treatments, such as bis(trifluoromethane) sulfonimide (TFSI) passivation, have also been used to increase the PLQY of MoS$_2$ from $\sim$0.8\% to over 95\% by reducing defect-mediated recombination~\cite{amani_near-unity_2015}. Strain engineering exploits spatially graded strain fields to funnel excitons toward regions of minimum bandgap, with reported PL intensity increasing a factor of $\sim$2 in MoS$_2$ and $\sim$2.8 in WSe$_2$~\cite{li_optoelectronic_2015, lee_drift-dominant_2022}. However, these approaches often rely on chemical treatments that degrade rapidly under ambient exposure~\cite{tanoh_giant_2021, li_mechanistic_2021}, or on uniform tensile strain that drives a direct-to-indirect bandgap transition, thereby suppressing rather than enhancing PL~\cite{conley_bandgap_2013}. Additionally, achieving high PL via hBN encapsulation typically requires vacuum annealing at $\sim$1000$^{\circ}$C~\cite{ryu_optical_2024}, a temperature typically incompatible with the use of pre-patterned electrodes, polymer substrates, and post-integration device architectures. These limitations motivate the need for PLQY control through non-invasive interlayer engineering. 

Forming vdW stacking enables tuning of electronic and optical properties through interlayer electronic coupling~\cite{geim_van_2013}, rather than solely through dielectric screening as in encapsulation~\cite{liu_van_2016}. In type-I heterostructures, both the conduction band minimum (CBM) and valence band maximum (VBM) of one layer lie within the gap of the other layer, confining electrons and holes to the same material and enhancing radiative recombination~\cite{bellus_type-i_2017}. In contrast, type-II heterostructures exhibit staggered band alignment, where the CBM and VBM reside in different layers, spatially separating electrons and holes and typically reducing the PL intensity. The PL enhancement in type-I heterostructures is typically attributed to exciton funneling, whereby carriers generated in the wider-bandgap layer transfer into the narrower-bandgap emissive layer and accumulate there~\cite{bellus_type-i_2017, altvater_efficient_2025}. This mechanism has been widely reported in systems combining an emissive TMD, such as MoSe$_2$ or WS$_2$, with a non- or weakly emissive material, for example, in FePS$_3$/MoSe$_2$ (showing a $\sim$20-fold enhancement of room-temperature PL emission from monolayer MoSe$_2$)~\cite{duan_enhanced_2021} and PbI$_2$/WS$_2$ (exhibiting a $\sim$2-fold PL enhancement relative to bare monolayer WS$_2$)~\cite{zheng_direct_2019} heterostructures. 

\begin{figure*}[htbp!]
\centerline{\includegraphics[width=180mm]{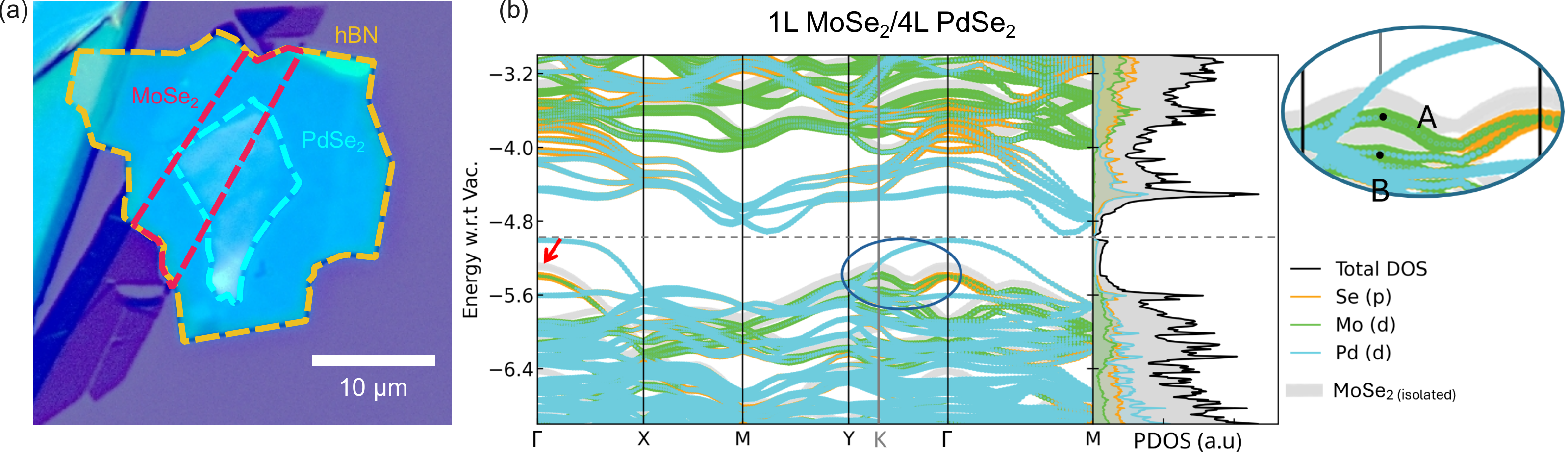}}
\caption{Structural and electronic properties of the MoSe$_2$/PdSe$_2$ heterostructure on an hBN substrate: (a) Optical image of the MoSe$_2/$PdSe$_2$/hBN heterostructure and (b) electronic band structures and projected density of states (PDOS) of a model heterostructure formed by 1L MoSe$_2$ and 4L PdSe$_2$. The vacuum level is set to 0~eV, and the Fermi level is marked with a horizontal dashed line in the mid-gap. The red arrow indicates the valence edges of 1L MoSe$_2$. The vertical bar between the high symmetry points Y and $\Gamma$ indicates the high symmetry point K of the unfolded primitive hexagonal lattice~\cite{recUC_MoTe2}. The inset shows the zoomed-in region with spin-split A- and B-exciton valleys of MoSe$_2$.}
\label{fig:Figure 1}
\end{figure*}

Interestingly, an unusual scenario has been demonstrated in a type-I heterostructure in which a strongly absorbing, non-luminescent narrow-gap SnSe$_2$ multilayer underlies a wider-gap monolayer MoS$_2$~\cite{dandu_strong_2019}. In that system, the MoS$_2$ PL is enhanced by up to 14-fold via a mechanism named Förster resonance energy transfer (FRET), that is, non-radiative transfer of the energy of photoexcited electron–hole pairs in SnSe$_2$ to the MoS$_2$ monolayer through near-field dipole–dipole coupling~\cite{dandu_strong_2019}. In type-I band alignments, one would expect carrier funneling into the narrower-gap layer, which should quench emission from the wider-bandgap TMD by depleting its photoexcited carriers. In this sense, MoS$_2$/SnSe$_2$ represents an unusual situation, as a strong interlayer energy transfer outweighs the PL loss pathways associated with charge transfer. This result established a general design principle: pairing a strongly absorbing, weakly emissive narrow-gap layer (SnSe$_2$) with a highly luminescent wider-gap TMD monolayer (MoS$_2$) can enhance, rather than quench, emission from the TMD. However, FRET efficiency in that system relies critically on the near-resonance between the direct transition of SnSe$_2$ and the excitonic gap of MoS$_2$, a condition that is not easily generalizable to other material combinations.

Here, we fabricate a vdW heterostructure combining a narrow-gap PdSe$_2$ multilayer with a monolayer of MoSe$_2$. PdSe$_2$ exhibits strong broadband absorption but weak PL~\cite{long_palladium_2019, slavich_multifunctional_2025}. Unlike SnSe$_2$, PdSe$_2$ does not exhibit a strong spectral resonance with the dominant excitonic transitions of MoSe$_2$, placing the system in a regime governed by interlayer electronic hybridization rather than dipole–dipole energy transfer. Its narrow-gap electronic structure and strong polarizability enable interactions beyond dielectric screening, in contrast to hBN or SiO$_2$. MoSe$_2$ is selected for its strong room-temperature emission~\cite{tongay_thermally_2012} and well-resolved A and B excitons ($\sim$200~meV splitting)~\cite{ross_electrical_2013, ugeda_giant_2014}, enabling sensitive tracking of exciton relaxation pathways~\cite{wang_colloquium_2018}. We investigate how interlayer coupling in a MoSe$_2$/PdSe$_2$/hBN vdW heterostructure modifies exciton relaxation and PLQY. We observe a pronounced enhancement of the MoSe$_2$ A-exciton PL, with the extracted PLQY increasing from $\sim$1\% to 6\%. 

Our approach exploits interlayer coupling in a vdW heterostructure, preserving intrinsic material properties. In addition to enhanced A-exciton emission, the heterostructure exhibits strong B-exciton quenching, which we attribute to interlayer hybridization of MoSe$_2$ with PdSe$_2$, due to the enabling of additional relaxation pathways. As a result, exciton populations are redistributed toward the lower-energy A-exciton, where radiative recombination is favored, leading to enhanced A-exciton emission. These findings demonstrate that PdSe$_2$ can serve as an effective platform for engineering exciton dynamics and enhancing PL efficiency in two-dimensional semiconductors without strain or chemical functionalization. Enhanced A-exciton PL in MoSe$_2$/PdSe$_2$ arises from interlayer coupling that requires near-resonant valence band alignment and low substrate density-of-states at the A-exciton energy, ensuring efficient radiative recombination.

\section{\label{Results}Results and Discussion}

We begin our analysis with the inspection of the optical image of a MoSe$_2$/PdSe$_2$/hBN heterostructure, which is shown in Figure \ref{fig:Figure 1}(a). The individual hBN, PdSe$_2$, and MoSe$_2$ flakes, as well as the overlapping MoSe$_2$/PdSe$_2$/hBN heterostructure area, are outlined by yellow dashed lines, light blue dashed lines, red dashed lines, and overlapping polygons, respectively. The thickness and layer quality of PdSe$_2$ and MoSe$_2$ were characterized by Raman spectroscopy and atomic force microscopy (AFM) in Figures \ref{fig:Figure S1} and \ref{fig:Figure S2}, respectively. The sample consists of a monolayer trigonal prismatic phase MoSe$_2$ flake partially overlapping a hexalayer (6L) PdSe$_2$ flake on an hBN substrate. At this thickness, PdSe$_2$ retains a finite bandgap of $\sim$1.0 eV while exhibiting strong broadband absorption that increases with layer number~\cite{lu_layer-dependent_2020}. In contrast, monolayer PdSe$_2$ has a larger bandgap ($\sim$1.3 eV)~\cite{oyedele_pdse2_2017, kim_quasiparticle_2021} and a weaker band offset with MoSe$_2$~\cite{zhang_direct_2014, zhang_band_2019}, limiting interlayer electronic coupling. On the other hand, thicker PdSe$_2$ flakes approach semimetallic behavior (bulk bandgap $\sim$0.03 eV)~\cite{oyedele_pdse2_2017}, and would possess a higher density of states which, in turn, could promote non-radiative recombination~\cite{liu_uncooled_2025}. In the following, we distinguish between two regions of the same MoSe$_2$ monolayer: (i) the heterostructure (HS) region, where MoSe$_2$ lies on PdSe$_2$, and (ii) the reference monolayer (1L) region, where MoSe$_2$ lies directly on hBN without PdSe$_2$ underneath. Both regions, therefore, contain 1L MoSe$_2$, but differ in the presence or absence of the PdSe$_2$ layer.

\begin{figure*}[htbp!]
\centerline{\includegraphics[width=180mm]{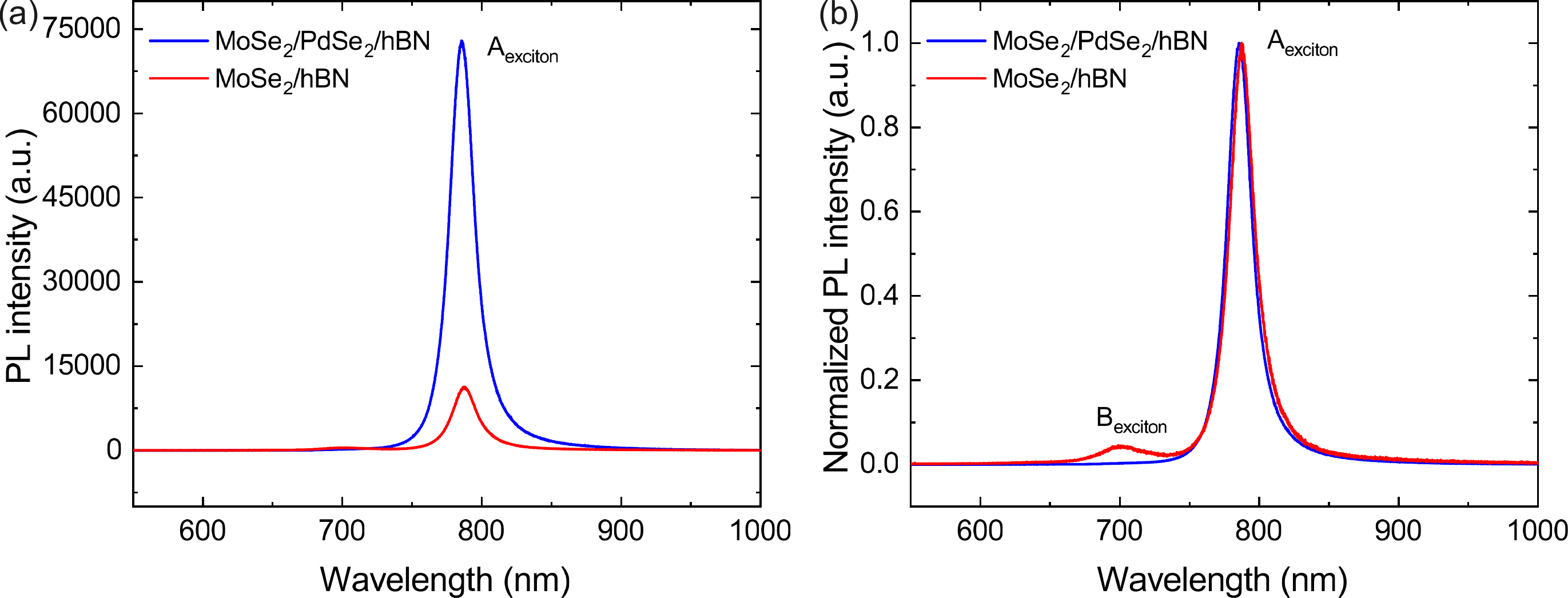}}
\caption{PL spectra of the 1L region and the HS region (a) A-exciton PL enhancement in the HS region at 514 nm excitation (b) B-exciton quenching in the HS region. PL intensity is reported in arbitrary units (a.u.), corresponding to detector counts.}
\label{fig:Figure 2}
\end{figure*}

To clarify the nature of interlayer electronic interactions in this system, we performed \textit{ab initio} simulations on a model system including a 1L MoSe$_2$ stacked on tetralayer (4L) PdSe$_2$ (Figure \ref{fig:Figure S3}), used as a computationally efficient approximation of the experimentally realized 6L PdSe$_2$ due to their similar electronic structures (Figure \ref{fig:Figure S4}). The electronic structure of 1L MoSe$_2$/4L PdSe$_2$ interface exhibits an almost vanishing indirect band gap, with both VBM and CBM dominated by PdSe$_2$ states (Figure~\ref{fig:Figure 1}(b)). The highest occupied states with MoSe$_2$ character are buried in the valence band of PdSe$_2$, between the VBM and the VBM-1 (the energy state located directly below the VBM) at the $\Gamma$ point. The interaction with the underlying PdSe$_2$ layer downshifts the uppermost valence bands of MoSe$_2$ by about 100 meV relative to the isolated 1L. Moreover, close to the K-valley of pristine MoSe$_2$ (inset of Figure~\ref{fig:Figure 1}(b)), where the A and B excitons originate, there are signatures of strong hybridization with PdSe$_2$ bands. This hybridization generates mixed electronic states that are expected to quench excitonic resonances in MoSe$_2$. Beyond opening new optical transition channels, this state mixing is anticipated to be particularly pronounced for the B-exciton, stemming from an even deeper valley interacting with the PdSe$_2$ valence region. Such interlayer coupling modifies the relaxation pathways of photoexcited carriers, enabling additional non-radiative interlayer relaxation channels, which compete against the intrinsic B~$\rightarrow$~A intralayer exciton relaxation. This alters exciton population dynamics, leading to a redistribution of carriers between A-exciton emission and interlayer decay channels, which may contribute to the enhanced A-exciton emission highlighted below. 

To validate this interpretation, we performed control calculations on an analogous 1L MoS$_2$/4L PdSe$_2$ HS (Figure \ref{fig:Figure S5}(a)). In this case, the valence states of MoS$_2$ are energetically deeper and do not hybridize as strongly with the valence band of PdSe$_2$. Consequently, exciton quenching is less pronounced, consistent with equivalent control measurements (Figures \ref{fig:Figure S5}(b,c)).

To investigate how band alignment affects excitonic emission, we performed PL measurements of the HS region. Figure \ref{fig:Figure 2}(a) shows a pronounced enhancement of the A-exciton PL in the HS region. Under 514 nm excitation, the A-exciton PL intensity increases by a factor $\sim$6× relative to the 1L region. Assuming comparable optical absorption at the excitation wavelength and identical light extraction efficiencies, we estimate an external PLQY of approximately 6\% for the HS, using the reported internal PLQY of 1\% for as-exfoliated 1L MoSe$_2$ as a benchmark~\cite{amani_recombination_2016}. Notably, while the absolute PL intensity of the ML is higher under 633 nm excitation due to its proximity to the A-exciton resonance, the relative enhancement factor decreases to $\sim$4× (Figure \ref{fig:Figure S6}). This suggests that the HS-induced enhancement is more pronounced under off-resonant excitation conditions. 

\begin{figure*}[htbp!]
\centerline{\includegraphics[width=150mm]{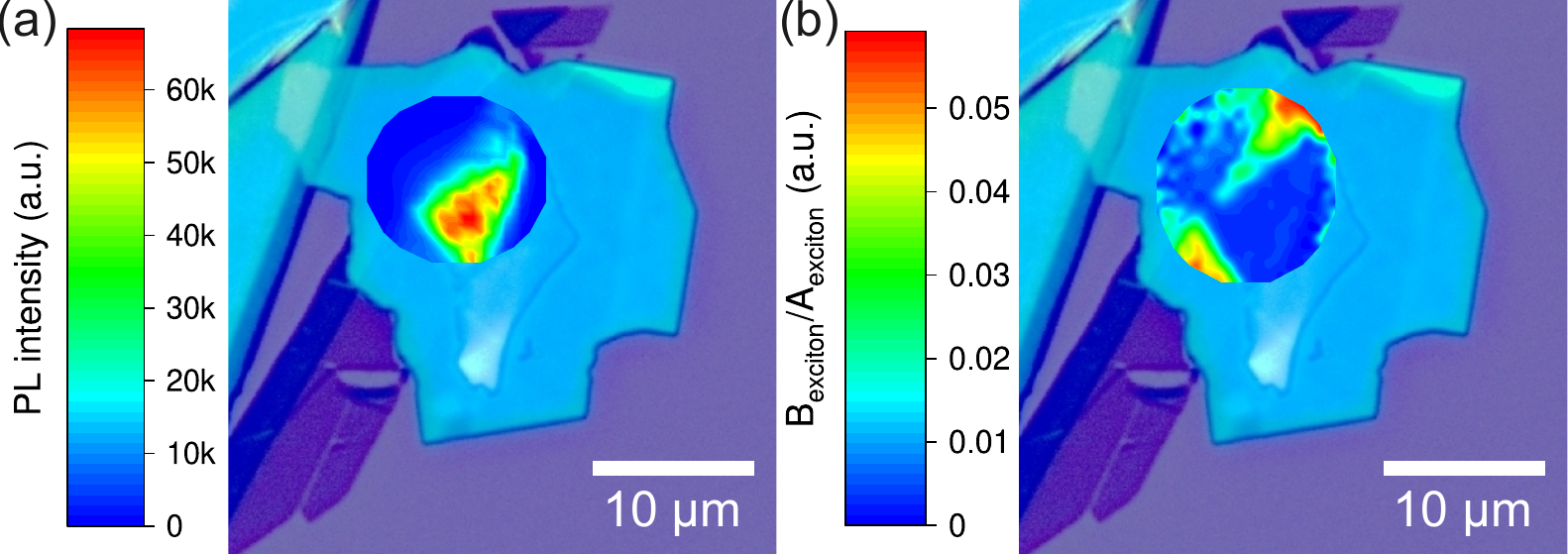}}
\caption{MoSe$_2$/PdSe$_2$ HS PL mapping a) A-exciton PL intensity map, I$_A$ (b) B/A exciton intensity ratio map, I$_B$/I$_A$}
\label{fig:Figure 3}
\end{figure*}

\begin{figure*}[htbp!]
\centerline{\includegraphics[width=180mm]{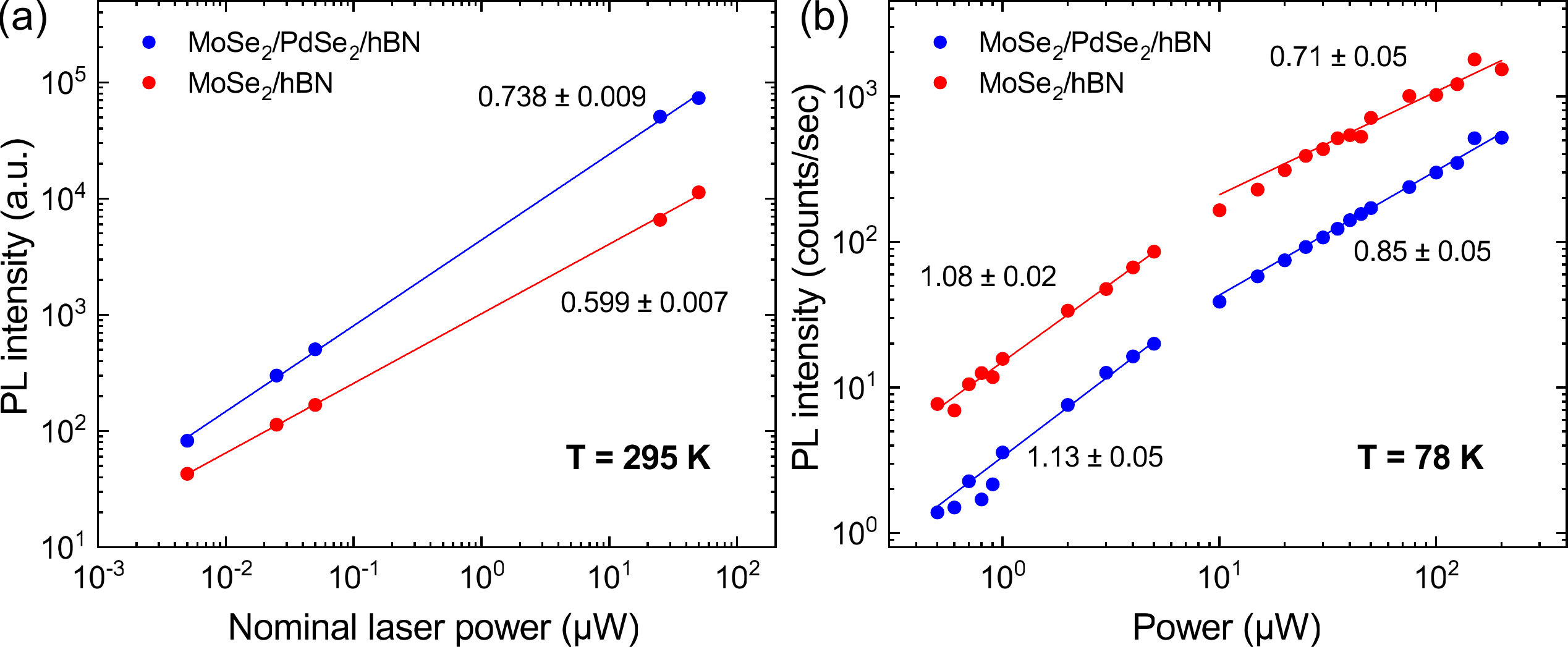}}
\caption{Laser power-dependent PL spectra of A-exciton at (a) room temperature and (b) cryogenic temperature. The reported power denotes the incident laser power at the sample.}
\label{fig:Figure 4}
\end{figure*}

In type-I TMD HSs, charge carriers transfer from the wider-bandgap layer to the narrower-bandgap emissive layer, resulting in enhanced radiative recombination~\cite{sun_band_2019, yamaoka_efficient_2018}. Although DFT calculations indicate a type-I band alignment for the MoSe$_2$/PdSe$_2$ HS (Figure \ref{fig:Figure 1}(b)), the observed PL enhancement occurs in the wider-bandgap MoSe$_2$. This contradicts the expected carrier funneling mechanism, as injection into the narrower-gap PdSe$_2$ would deplete the MoSe$_2$ exciton population and suppress emission. Similarly, while dielectric screening from a dielectric substrate like hBN can influence the excitonic behavior of TMDs, it fails to account for the PL enhancement observed in this HS. The specificity of this effect is underscored by the fact that PdSe$_2$ does not enhance PL in MoS$_2$ (Figures \ref{fig:Figure S5}(b,c)), ruling out generic substrate-induced passivation. Instead, we attribute the enhancement to specific electronic interactions between PdSe$_2$ and MoSe$_2$. Supporting this interpretation, Figure \ref{fig:Figure 2}(b) reveals a strong quenching of the B-exciton emission in the HS region, consistent with the interlayer hybridization predicted by our DFT calculations. Furthermore, the minimal blueshift ($\sim 3$ meV)~\cite{varjamo_optical_2025} and the absence of a clear trion (charged exciton) enhancement~\cite{lone_manipulation_2025} suggest that charge-transfer-induced doping is negligible.  Consequently, neither simple carrier funneling nor doping variations can be considered the dominant drivers of the observed PL intensity modification. 

To prove the robustness of our analysis, we examined the consistency of this behavior across the HS. As shown in Figure \ref{fig:Figure 3}(a), the PL emission is uniform across the entire HS region, demonstrating the spatio-spectral homogeneity of the A-exciton PL. The PL mapping in Figure \ref{fig:Figure 3}(b) further confirms low B-exciton population throughout the HS area. 

To further probe the underlying mechanisms of the observed A-exciton enhancement and B-exciton quenching, we investigated the dependence of the PL intensity on excitation power and temperature. The data presented in Figure \ref{fig:Figure 4} are well described by a power-law dependence (I $\propto$ P$^\alpha$). As shown in Figure \ref{fig:Figure 4}(a), the PL intensity of both the 1L region and the HS region increases sublinearly with excitation power, yielding exponents of $\alpha$ $\simeq$ 0.599 and $\alpha$ $\simeq$ 0.738, respectively. An exponent significantly smaller than unity ($\alpha < 1$) in the 1L region typically indicates that relaxation dynamics are dominated by exciton-exciton annihilation (EEA), a process whereby two excitons interact, and one or both recombine non-radiatively, reducing the PL intensity at high excitation densities~\cite{yuan_exciton_2015}. The increase of the exponent to $\alpha$ $\simeq$ 0.738 in the HS region suggests a suppression of EEA, consistent with interlayer hybridization modifying exciton relaxation pathways. At cryogenic temperatures, Figure \ref{fig:Figure 4}(b) reveals two distinct regimes. In the low-power region, the PL intensity scales nearly linearly with excitation power, yielding exponents of $\alpha$ $\simeq$ 1.08 for the 1L region and $\alpha$ $\simeq$ 1.13 for the HS region. At higher excitation powers, the response becomes sublinear, with $\alpha$ $\simeq$ 0.71 and $\alpha$ $\simeq$ 0.85, respectively, which is attributed to EEA at high carrier densities~\cite{cai_ultralow_2024}. Importantly, unlike the room-temperature behavior, the large PL enhancement disappears at 78 K, with the HS region exhibiting lower intensity than the 1L region. This observation indicates that the mechanisms responsible for the A-exciton enhancement are strongly temperature-dependent, consistent with a temperature-dependent redistribution of exciton relaxation dynamics that favors radiative recombination at intermediate temperatures ($\sim$200-300 K) and weakens at low temperatures (< 200 K).

\begin{figure*}[htbp!]
\centerline{\includegraphics[width=180mm]{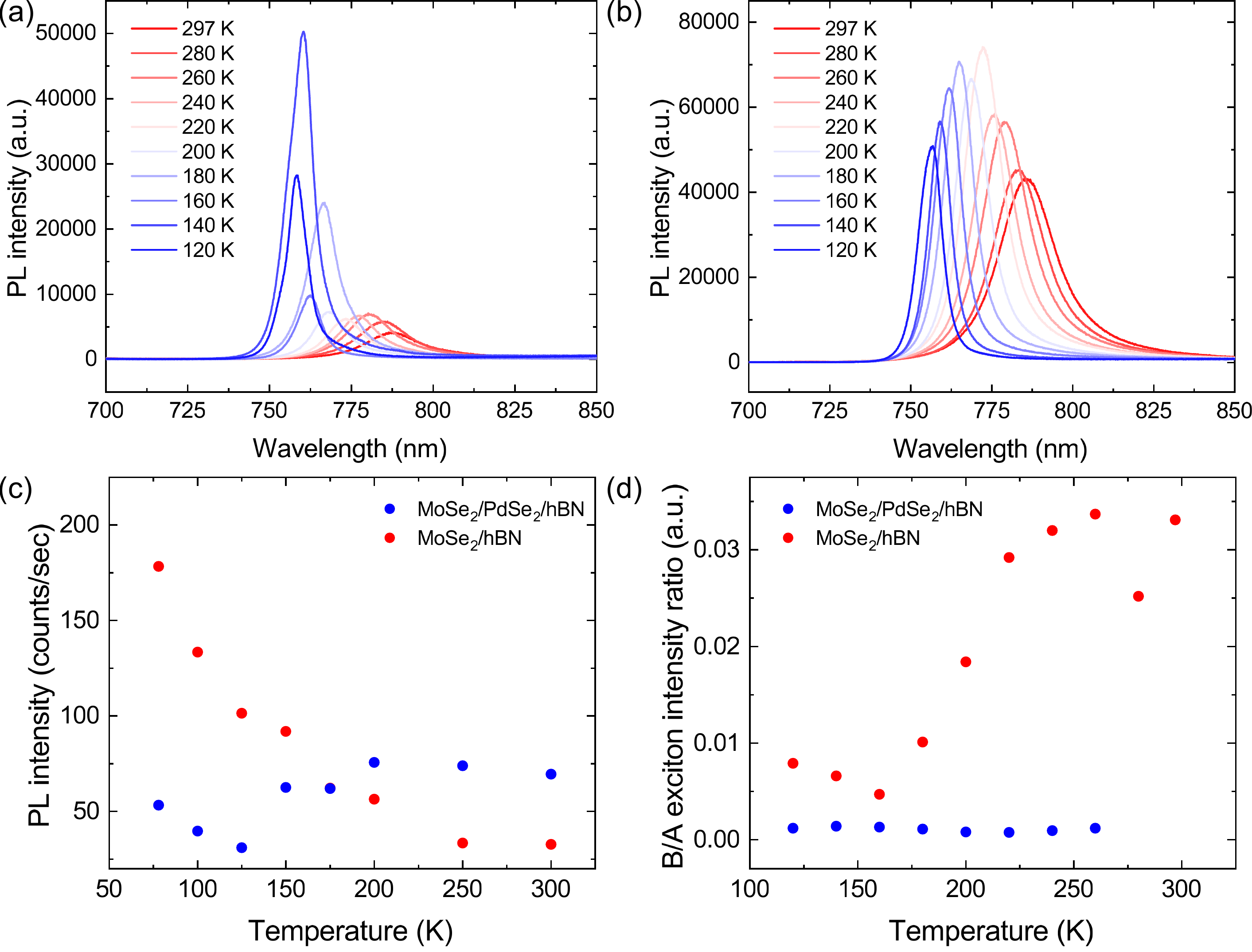}}
\caption{Temperature-dependent PL measurements of (a) the 1L region and (b) the HS region (c) Temperature-dependent peak PL measurements of the 1L region and the HS region (d) Temperature dependence of the 1L region and the HS region B-exciton-to-A-exciton intensity ratio.}
\label{fig:Figure 5}
\end{figure*}

Figure \ref{fig:Figure 5} presents temperature-dependent PL measurements of the HS region and the 1L region. The 1L region exhibits a conventional thermal quenching behaviour~\cite{tongay_thermally_2012}, with the PL intensity decreasing as temperature increases (Figure \ref{fig:Figure 5}(a)). This behavior is attributed to the activation of non-radiative recombination processes~\cite{tongay_thermally_2012}, as exciton–phonon interactions at elevated temperatures provide sufficient thermal energy for excitons to access non-radiative pathways. Concurrently, the PL peak energy redshifts with increasing temperature, consistent with lattice thermal expansion~\cite{tongay_thermally_2012}. In contrast, the HS region displays an anomalous temperature dependence. As shown in Figure \ref{fig:Figure 5}(b), the PL intensity exhibits a non-monotonic trend: it initially increases upon cooling from room temperature, reaching a maximum at approximately 220 K, and subsequently decreases upon further cooling to 120 K.  By comparison, the 1L A-exciton PL increases monotonically with decreasing temperature (Figure \ref{fig:Figure 5}(c)). This contrasting behavior demonstrates that PdSe$_2$ modifies exciton relaxation dynamics in MoSe$_2$. The reduced PL intensity in the HS region below $\sim$200 K reflects a temperature-dependent change in the balance of radiative and non-radiative recombination processes. This behavior is consistent with reduced phonon-assisted scattering~\cite{wang_colloquium_2018}, which can limit dark–bright exciton conversion and intervalley relaxation~\cite{brem_phonon-assisted_2020, selig_excitonic_2016}, thereby reducing the repopulation of the radiative A-exciton state. This behavior is hence not governed by EEA effects. Figure \ref{fig:Figure 5}(d) shows the temperature-dependent intensity ratio of the B-exciton to the A-exciton. In the 1L region, the ratio exhibits an overall increasing trend, reaching approximately 0.03 at 300 K, consistent with exciton thermalization governed by Boltzmann statistics~\cite{wang_colloquium_2018}. In contrast, the HS region exhibits a nearly temperature-independent B/A ratio, suggesting modified exciton thermalization dynamics in the presence of the PdSe$_2$ interface.

\begin{figure*}[htbp!]
\centerline{\includegraphics[width=180mm]{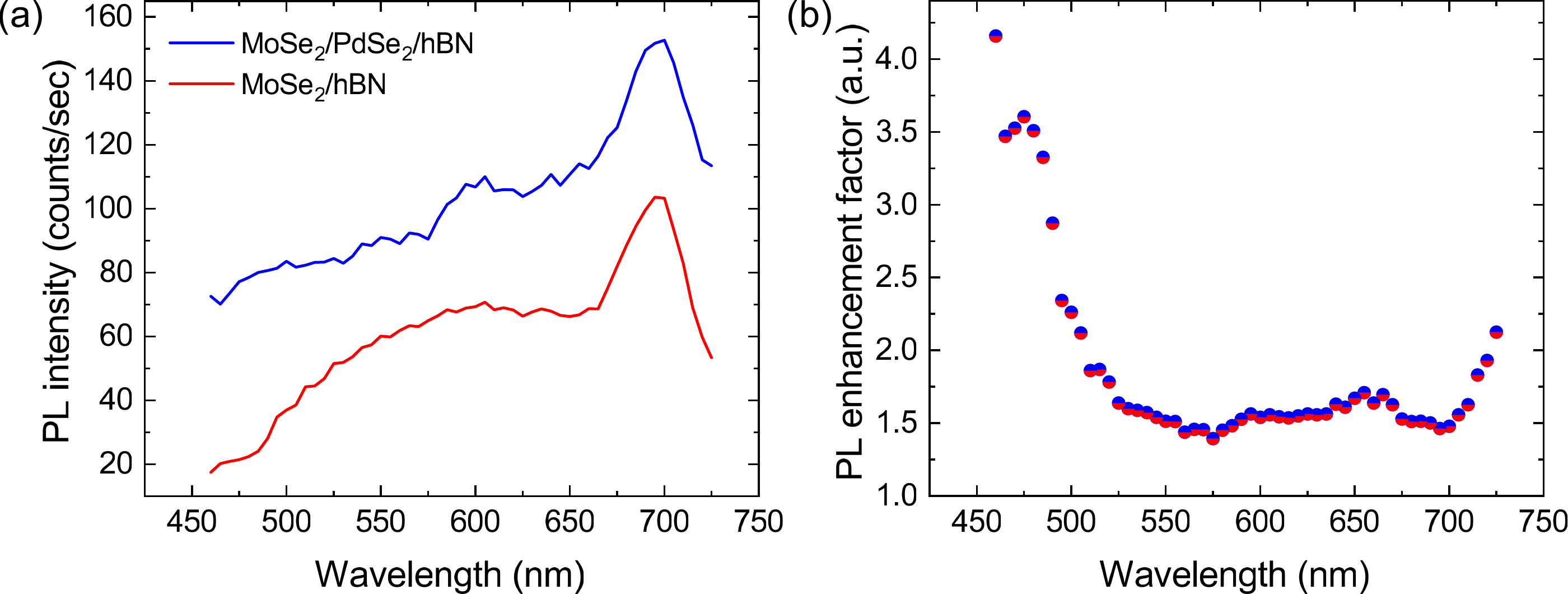}}
\caption{(a) PLE spectra of the 1L and the HS (b) PLE intensity ratio between the HS and the 1L.}
\label{fig:Figure 6}
\end{figure*}

To further probe how interlayer interactions affect exciton generation and recombination, we performed photoluminescence excitation (PLE) spectroscopy on the HS. Unlike conventional PL measurements, PLE examines how the excitation energy influences the emission intensity, providing insight into absorption pathways, exciton relaxation, and interlayer coupling~\cite{estrada-real_probing_2023}. PLE spectroscopy reveals a consistent broadband PL enhancement of emission from the HS region across the entire excitation energy range compared to the 1L region (Figure \ref{fig:Figure 6}(a)). Although excitation near 700 nm, corresponding to the MoSe$_2$ B-exciton resonance, produces the highest absolute PL intensity due to stronger intrinsic absorption, the PL enhancement persists across the full spectral range ($\sim$450–725 nm) rather than being confined to a specific transition.
The wavelength-dependent PL enhancement factor extracted from the PLE spectra (Figure \ref{fig:Figure 6}(b)) reveals the largest PL enhancement at shorter excitation wavelengths, exceeding $\sim3.5$ near 450 nm, while decreasing to approximately $\sim1.5$ in the $\sim$550–650 nm range but remaining greater than unity. Unlike the PL quenching across the whole range observed in MoS$_2$/PdSe$_2$ (Figure \ref{fig:Figure S8}), the MoSe$_2$-based HS exhibits a divergent excitation response, demonstrating that interfacial band alignment dictates the transition between emission enhancement and suppression. 

\section{\label{End}Conclusions}

We have shown that interlayer coupling in MoSe$_2$/PdSe$_2$ HSs enables strong enhancement of A-exciton emission in 1L MoSe$_2$, increasing the PLQY from $\sim$1\% in the reference monolayer to 6\% at room temperature. Our method exploits interlayer coupling in a fully assembled heterostructure, enhancing emission while preserving the intrinsic properties of the monolayer. The enhancement is accompanied by pronounced B-exciton quenching, reduced EEA, and a broadband increase in excitation efficiency. The observation that emission is enhanced in the wider-bandgap layer, together with the broadband PLE response and temperature-dependent crossover to quenching, indicates that the effect does not arise from simple exciton funneling or a single resonant process. Instead, the combined PL, broadband excitation PLE measurements, temperature-dependent PL, and DFT calculations indicate a mechanism in which interlayer electronic interactions redistribute exciton populations and modify relaxation pathways. 

These findings establish interlayer coupling as a means to control exciton dynamics and enhance radiative recombination in two-dimensional semiconductors, offering a general strategy for improving light emission in vdW HSs. The enhanced A-exciton PL in MoSe$_2$/PdSe$_2$ arises from favorable interlayer coupling, which requires two criteria: (i) the monolayer VBM lies near a high-density-of-states region of the narrow-gap substrate, enabling hybridization (the $\sim$100 meV offset in MoSe$_2$/PdSe$_2$ meets this, while MoS$_2$ does not), and (ii) the substrate has low density-of-states at the A-exciton energy, preventing excessive transfer and allowing radiative recombination to dominate. Because these criteria depend on band alignment and density of states, which can be tuned through material selection and stacking in a wide range of vdW HSs, they open opportunities for systematic design of materials with controlled exciton dynamics and improved light emission.

\raggedbottom
\newpage

\subsection{\label{Sec:Exp}Experimental Methods}

\textit{Crystal growth}: 1T-PdSe$_2$ was made by flux growth in a quartz ampoule using Pd and Se. The stoichiometric amount of Pd (powder sponge, 99.95\%, Safina, Czech Republic) and selenium (2-4 mm granules, 99.9999\%, Wuhan Xinrong New Materials Co., China) were placed in stoichiometric amount corresponding to 3 g in quartz ampoule (25 x 100 mm$^{2}$ and melt sealed by oxygen-hydrogen welding torch under high vacuum (<1$\cdot$10$^{-3}$ Pa, oil diffusion pump with LN$_2$ trap). The ampoule was placed in a muffle furnace and heated at 850 °C (heating rate 1 °C/min), and after 5 hours, cooled to 500 °C using a cooling rate of 6 °C/hour and finally freely cooled to room temperature. Ampoules were opened in an argon-filled glovebox, and crystals were collected.
2H-MoSe$_2$ was made by the chemical vapor transport (CVT) method using bromine as a transport medium. Selenium (2-4 mm granules, 99.9999\%, Wuhan Xinrong New Materials Co., China) and molybdenum (-100 mesh, 99.95\%, Molymet, Germany) were placed in ampoule (40 x 250 mm$^{2}$) with stochiometric ratio corresponding to 30 g of MoSe$_2$ together with 1 g of SeBr$_4$ (99\%, Strem, USA) and melt sealed by oxygen-hydrogen welding torch under high vacuum (<1$\cdot$10$^{-3}$ Pa, oil diffusion pump with LN$_2$ trap). The ampoule was placed in a muffle furnace and heated at 500 °C for 25 hours, at 600 °C for 50 hours, and finally at 800 °C for 50 hours. The heating and cooling rates were 1 °C/min. The ampoule was placed in a two-zone furnace. First, the growth zone was heated to 1000 °C and the source zone to 800 °C, and after 2 days, the thermal gradient was reversed, and the growth zone was heated to 900 °C and the source zone to 1000 °C for 14 days. Finally, the ampoule was cooled to room temperature and opened in an argon-filled glovebox.

\textit{HS preparation}: 1L MoSe$_2$ (and MoS$_2$), multilayer PdSe$_2$, and hBN flakes were mechanically exfoliated from their respective bulk crystals. The HSs MoSe$_2$/PdSe$_2$/hBN (and MoS$_2$/PdSe$_2$/hBN) were fabricated on SiO$_2$/Si substrates using a standard dry transfer method with a polydimethylsiloxane (PDMS) stamp~\cite{castellanos-gomez_deterministic_2014}. Using an optical microscope and an aligned manual transfer system (HQ Graphene), the layers were successively transferred onto the substrate in the following sequence: bottom hBN, PdSe$_2$ multilayer, and finally the MoSe$_2$ (or MoS$_2$) 1L.

\textit{AFM characterization}: To probe the surface topography and determine the layer thickness, atomic force microscopy (AFM) measurements were performed at room temperature using a Bruker Dimension Icon instrument (Bruker Corporation, Billerica, MA, USA) and the ScanAsyst-Air cantilever (Bruker Corporation, Billerica, MA, USA) with a resonance frequency of 70 kHz, a spring constant of 0.4 N/m, and a nominal tip radius of 2 nm. Height images of the HS region, the 1L region, and PdSe$_2$ flakes were acquired using the PeakForce Tapping mode, operating at the off-resonance frequency of 2 kHz. Force set-point and scan rate were set to 0.5 nN and 0.6 Hz, respectively.

\textit{Computational methods}: All calculations were performed within the framework of DFT~\cite{dft_Hohenberg1964} using the Vienna ab-initio simulation Package (VASP)~\cite{vasp_Kresse1996}. The electron-ion interactions in the Kohn-Sham equations~\cite{Kohn-sham1965} were treated with the projector augmented-wave method~\cite{PAW_method}, and the exchange-correlational effects were described with the Perdew, Burke, and Ernzerhof functional~\cite{GGA_Perdew1996} within the generalized gradient approximation. vdW interactions were included through the optPBE-vdW density-functional scheme~\cite{vdw_DF}, which provides the most accurate description of the lattice parameters of multilayer PdSe$_2$~\cite{PdSe2-2D}. A plane wave energy cutoff of 500~eV was employed, and the Brillouin zone was sampled using a $12 \times 12 \times 1$ $k$-point mesh. Spin-orbit coupling was included in all electronic structure calculations; in these cases, the increased computational cost was mitigated by reducing the energy cutoff to the \texttt{ENMAX} value of the employed pseudopotentials. The self-consistent field and structural relaxation were converged to $10^{-6}$~eV for the total energy and 15~meV~\AA$^{-1}$ for the residual forces. To eliminate artificial interactions between periodic images, a 16~\AA\ vacuum region was introduced perpendicular to the 1L plane. The crystal structures were visualized using \texttt{ASE}~\cite{ASE_tool}, and the post-processing code \texttt{SUMO}~\cite{sumo-plot} was used to plot the band structures and the PDOS.

\textit{Photoluminescence and Raman measurements}: PL and Raman measurements were performed using a Renishaw inVia confocal Raman microscope (Renishaw, UK). High-frequency Raman spectra were acquired with a 514 nm excitation laser at 0.05 mW through a 100$\times$ objective, dispersed by a 1800 lines/mm holographic grating. Low-frequency Raman measurements used the same laser at 0.05 mW, with an ECLIPSE filter and a 2400 lines/mm grating, 300 s exposure, and a 100$\times$ objective. PL spectra were recorded using 514 nm or 633 nm excitation at 0.05 mW and 0.075 mW, dispersed by a 1200 lines/mm grating with 10 s exposure through a 100$\times$ objective. PL mapping employed a 514 nm laser at 0.5 mW with a 1200 lines/mm grating, 10 s exposure per pixel, and a 100$\times$ objective, with spatial maps constructed from the A-exciton peak intensity at 786 nm and the B/A exciton intensity ratio (702/786 nm). Temperature-dependent PL was performed using a Linkam HFS600E-PB4 heating/cooling stage, exciting the sample with a 514 nm laser (0.05 mW) through a long-working-distance 50$\times$ objective, dispersed by a 1200 lines/mm grating with 10 s exposure. Power-dependent PL measurements at room temperature were performed using the same Renishaw inVia confocal Raman microscope by varying the excitation power of the 514 nm laser. Power-dependent PL measurements at cryogenic temperatures, along with temperature-dependent PL studies, were performed using a custom-built micro-spectroscopy setup that includes a continuous-flow cryostat (Janis ST-500). The emitted signal was dispersed by a Horiba iHR-320 monochromator and detected using a thermoelectrically cooled CCD camera (Syncerity).

\textit{PLE}: Room-temperature (295 K) photoluminescence excitation (PLE) measurements of the hBN/MoSe$_2$ and MoSe$_2$/PdSe$_2$/hBN HSs were performed using a custom-built micro-spectroscopy setup. The samples were mounted on a Thorlabs PT3 XYZ mechanical stage using silver paste to ensure position control and mechanical stability. Excitation was provided by a tunable supercontinuum white-light laser (NKT Photonics) and focused onto the sample through a Mitutoyo 50$\times$/0.42 NA objective, resulting in a spot size of approximately 1\,$\mu$m. The excitation wavelength was tuned from 460 nm to 725 nm in steps of 5 nm, while the incident laser power was maintained constant at 10\,$\mu$W across the entire spectral range. To suppress laser scattering, a 775 nm short-pass filter was used in the excitation path, while a corresponding 775 nm long-pass filter was placed in the collection path (both from Edmund Optics). All spectra were acquired with an exposure time of 5\,s and a single accumulation for each excitation wavelength.

\clearpage
\widetext

\setcounter{equation}{0}
\setcounter{figure}{0}
\setcounter{table}{0}
\setcounter{page}{1}
\makeatletter
\renewcommand{\theequation}{S\arabic{equation}}
\renewcommand{\thefigure}{S\arabic{figure}}
\renewcommand{\thetable}{S\arabic{table}}
\renewcommand{\bibnumfmt}[1]{[S#1]}
\renewcommand{\citenumfont}[1]{S#1}
\newcounter{SIfig}
\renewcommand{\theSIfig}{S\arabic{SIfig}}

\section*{Supporting Information}

\subsection{Raman spectroscopy}

\begin{figure*}[htbp!]
\centerline{\includegraphics[width=80mm]{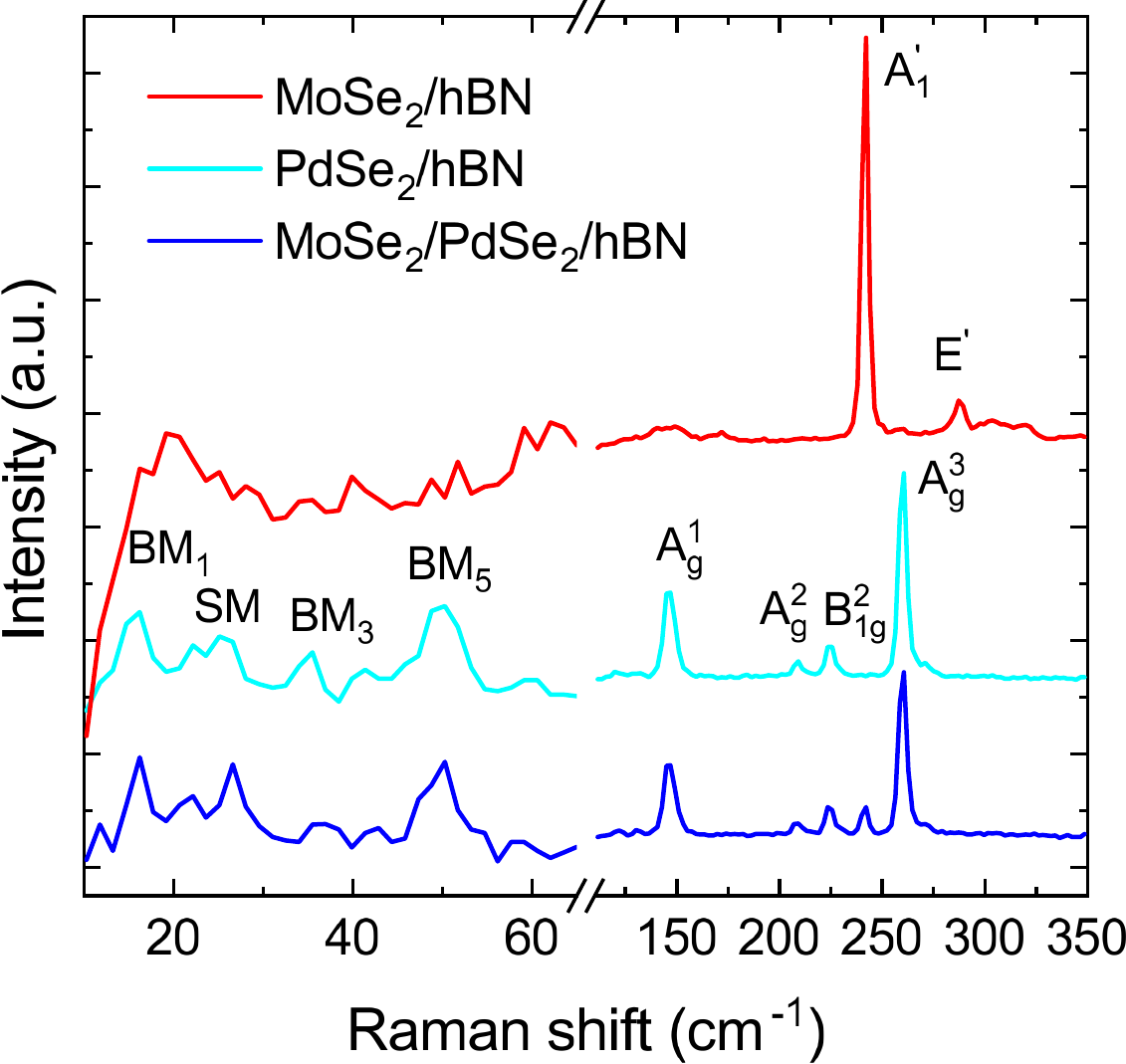}}
\caption{Raman spectra acquired from the 1L region, PdSe$_2$ on hBN, and the HS region}
\label{fig:Figure S1}
\end{figure*}

Figure \ref{fig:Figure S1} presents the Raman spectra acquired from three regions: PdSe$_2$ on hBN, the 1L region, and the HS region. Low-frequency Raman spectroscopy (0–100 cm$^{-1}$) is a widely used, non-destructive probe of interlayer vibrational modes in two-dimensional materials. For PdSe$_2$, the corrected linear chain model~\cite{puretzky_anomalous_2018} can be used to estimate the number of layers from the peak positions of shear (SM) and layer-breathing (BM) modes via

\begin{equation}
\begin{aligned}
\omega(B_j) \approx \sqrt{2}\,\omega(L, B_1)(1 - \beta)
\left[ \sin^2\!\left( \frac{j\pi}{2N} \right)
+ \beta \sin^2\!\left( \frac{j\pi}{N} \right) \right]^{1/2}
\end{aligned}
\end{equation}

where $\omega(B_j)$ is the Raman shift (in cm$^{-1}$) of the j-th BM branch $\omega(L, B_1)$=43.2 cm$^{-1}$, $\beta$=0.13, and $N$ is the number of layers. From this analysis, the extracted layer number of PdSe$_2$ is approximately six layers. The identification of the 1L region is supported by the absence of low-frequency Raman peaks, as these modes originate from interlayer vibrations. In the high-frequency range (100-350 cm$^{-1}$), the 1L region exhibits two characteristic Raman peaks corresponding to the A$^{'}_{1}$ (242 cm$^{-1}$) and E$^{'}$ (287 cm$^{-1}$) modes. In contrast, PdSe$_2$ shows four distinct peaks associated with the A$^{1}_{g}$ (146 cm$^{-1}$), A$^{2}_{g}$ (209 cm$^{-1}$), B$^{2}_{1g}$ (225 cm$^{-1}$), and A$^{3}_{g}$ (261 cm$^{-1}$). In the HS region, all PdSe$_2$ peaks remain clearly visible, whereas the Raman intensities of the MoSe$_2$ modes are suppressed, indicating an interaction between the two layers.

\raggedbottom
\newpage

\subsection{\label{SI_AFM}AFM topography}

\begin{figure*}[htbp!]
\centerline{\includegraphics[width=180mm]{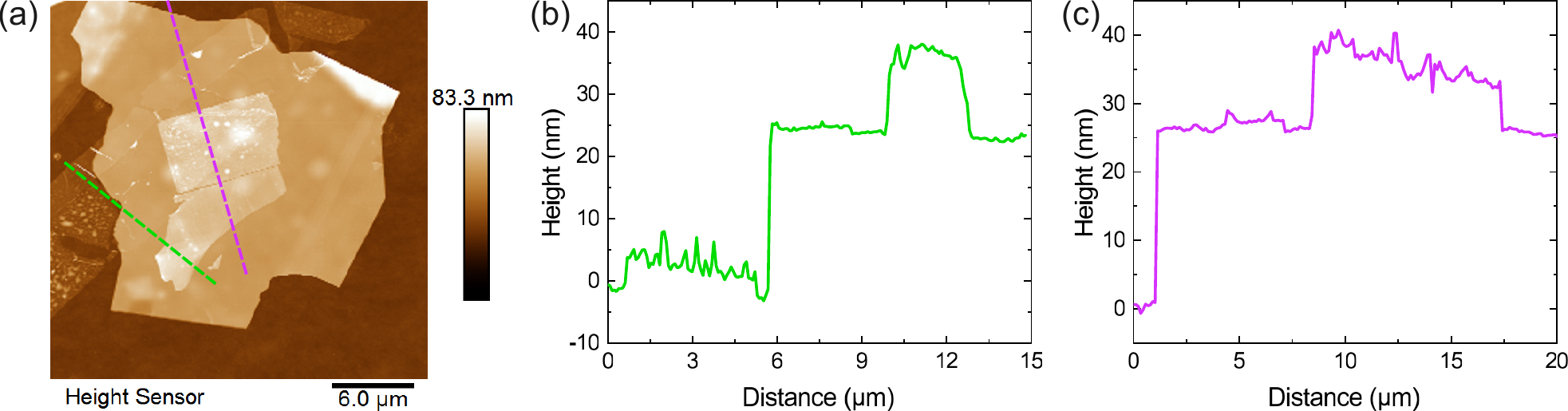}}
\caption{(a) AFM topography of the measured area of the MoSe$_2$/PdSe$_2$ HS on hBN showing regions with different thicknesses; (b, c) corresponding height profiles taken along the green and magenta dashed lines in (a)}
\label{fig:Figure S2}
\end{figure*}

AFM was used to verify the thickness and vertical stacking of the MoSe$_2$/PdSe$_2$ HS on hBN. The topography map of the measured region is shown in Figure \ref{fig:Figure S2}(a), where terraces corresponding to the different constituent layers are clearly visible. The height profile extracted along the dashed line in Figure \ref{fig:Figure S2}(b) traverses a region without overlap between PdSe$_2$ and MoSe$_2$, providing an accurate measurement of the 1L MoSe$_2$ thickness. Along the scan, the first terrace corresponds to 1L MoSe$_2$ on hBN, with a step height of $\sim$0.8 nm above the hBN surface. This is followed by the bare hBN region ($\sim$22.4 nm), forming the lowest terrace, and then a multilayer PdSe$_2$ region ($\sim$7.7 nm above hBN). The profile then returns to the hBN-only baseline. This scan is therefore useful for precisely quantifying the 1L MoSe$_2$ thickness and verifying the PdSe$_2$ step height, confirming the vertical dimensions of the individual constituents.

A separate line profile along the path shown in Figure \ref{fig:Figure S2}c traverses the PdSe$_2$ region and captures the HS. The lowest terrace corresponds to bare hBN ($\sim$26.3 nm), followed by a multilayer MoSe$_2$ region on hBN ($\sim$27.7 nm) and a 1L MoSe$_2$ on hBN ($\sim$26.4 nm). The overlapping HS region, where MoSe$_2$ sits on top of PdSe$_2$, exhibits a height of $\sim$37.9 nm, consistent with the sum of the individual layers, and displays a relatively smooth surface with an RMS roughness below 1.4 nm. The next terrace corresponds to the PdSe$_2$ multilayer on hBN ($\sim$34.1 nm). Notably, low-frequency Raman measurements indicate that this PdSe$_2$ flake consists of six layers, whereas the AFM-measured step height of $\sim$7.7–8.5 nm is slightly larger than the expected crystallographic thickness for six layers, a discrepancy commonly observed in AFM measurements of 2D materials due to tip-sample interactions, adsorbates, or surface roughness~\cite{plougmann_how_2021, purckhauer_analysis_2019}. The scan then returns to the bare hBN baseline ($\sim$25.6 nm). Together, these profiles confirm the vertical stacking hierarchy of the HS and provide direct thickness measurements consistent with Raman-based layer assignment.

\clearpage
\widetext

\subsection{Lattice parameters optimization}

\begin{figure*}[htbp!]
\centerline{\includegraphics[width=75mm]{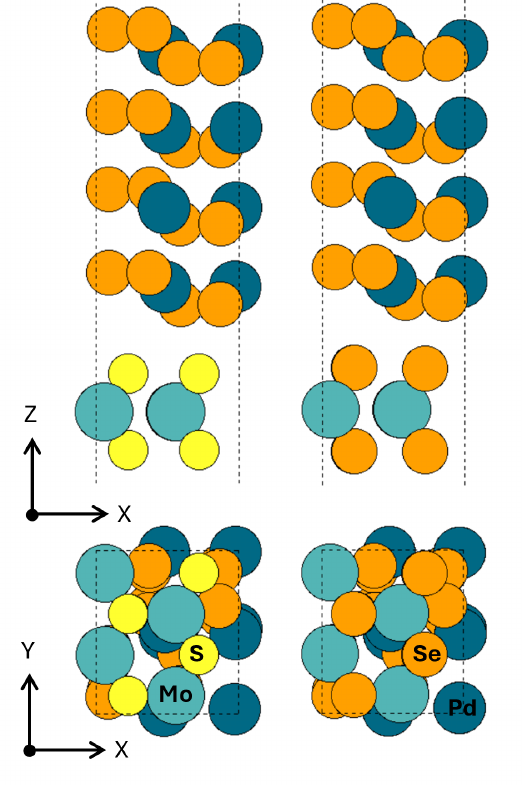}}
\caption{Top and side view of the crystal structure of 1L MoS$_2$ and 1L MoSe$_2$ stacked on 4L PdSe$_2$}
\label{fig:Figure S3}
\end{figure*}

In our simulations, the 4L PdSe$_2$ and the 1L MoS$_2$ and 1L MoSe$_2$ were taken from their orthorhombic and 2H bulk phases, respectively. The relaxed in-plane lattice parameters of the pure layers are: a = 5.820 Å, b = 5.974 Å for 4L PdSe$_2$, a = b = 3.166 Å for 1L MoS$_2$, and a = b = 3.298 Å for 1L MoSe$_2$, in good agreement with the previous reports~\cite{PdSe2-2D,tmdc_hs_ramzan,direct-ID-MoS2}, confirming the accuracy of our adopted simulation parameters. The HSs were built by stacking a 2×2 rectangular supercell of MoS$_2$ (or MoSe$_2$) on a 1×1 cell of 4L PdSe$_2$ unit cell, resulting in 36 atoms per simulation cell, as shown in Figure \ref{fig:Figure S3}.
The in-plane lattice vectors and atomic positions were fully relaxed using vdW-corrected DFT (see Computational methods) to diffuse the lattice mismatch between layers. The final relaxed lattice parameters are a = 5.842 Å and b = 6.250 Å for 1L MoSe$_2$/4L PdSe$_2$, and a = 5.721 Å and b = 6.132 Å for 1L MoS$_2$/4L PdSe$_2$. The resulting interlayer distance between the topmost Se atom of 4L PdSe$_2$ and the nearest Se and S atoms of the 1L MoSe$_2$ and 1L MoS$_2$ is 3.11 Å and 3.04 Å, respectively. These vdW gaps obtained after full relaxation are within the typical range reported for TMD-based HSs\cite{tmdc_hs_ramzan}.

\clearpage
\widetext

\subsection{\label{DFT}Electronic structure calculations}

\begin{figure*}[htbp!]
\centerline{\includegraphics[width=100mm]{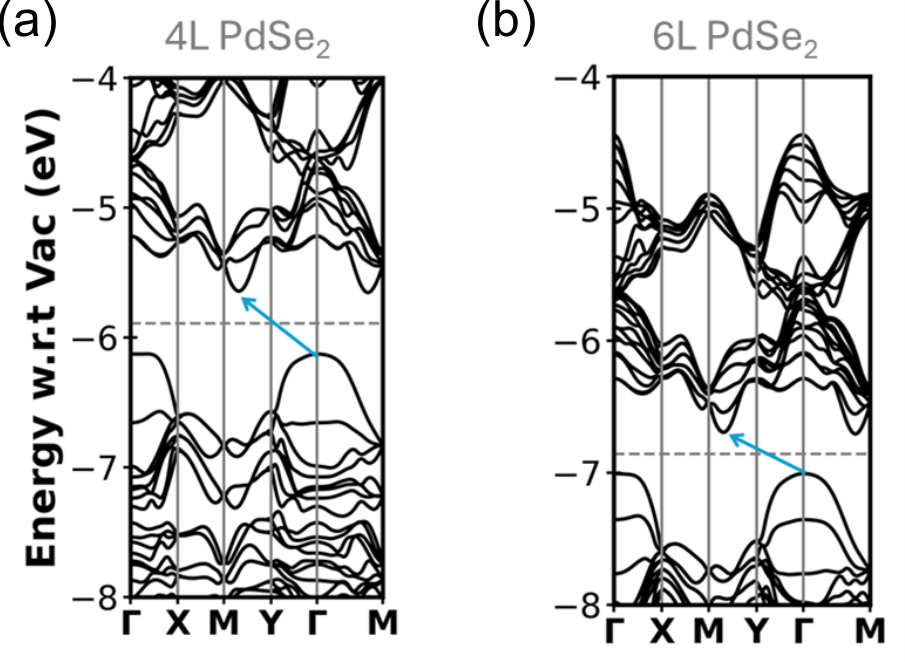}}
\caption{Electronic band structure of (a) 4L- and (b) 6L-PdSe$_2$ calculated with opt-PBE vdW functionals. The vacuum level is set to zero, and the blue arrows mark fundamental gaps. The Fermi-level marked with a gray-dashed line is set to the mid band gap.}
\label{fig:Figure S4}
\end{figure*}

To investigate the influence of the vdW interface on the electronic properties, \textit{ab initio} simulations were performed. As shown in Figures \ref{fig:Figure S4}(a,b), the calculated electronic band structures of 4L and 6L PdSe$_2$ exhibit qualitatively analogous electronic features. Consequently, 4L PdSe$_2$ was utilized as a computationally efficient proxy for the experimental 6L PdSe$_2$ in the HS simulations.

\begin{figure*}[htbp!]
\centerline{\includegraphics[width=180mm]{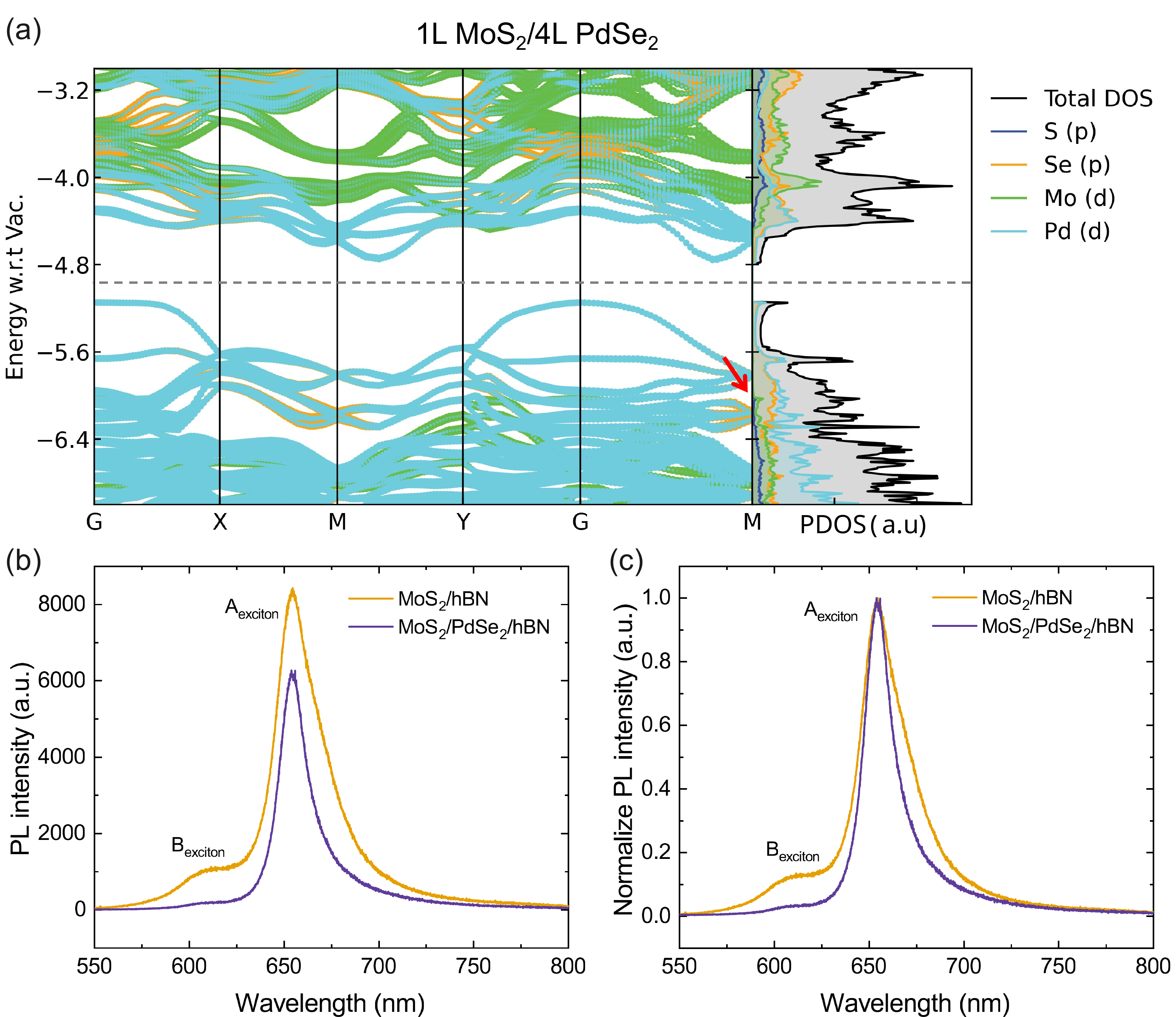}}
\caption{(a) Electronic band structures alongside atom PDOS of 1L MoS$_2$/4L PdSe$_2$. The vacuum level is set to zero, and the Fermi level marked with a horizontal dashed line is set to the midgap. The vertical grey line between high symmetry points Y and $\Gamma$ corresponds to the folded high symmetry point K of the primitive hexagonal lattice. PL spectra of the 1L region and the HS region of MoS$_2$-based samples at 514 nm excitation (b) A-exciton quenching in the HS region (c) B-exciton quenching in the HS region.}
\label{fig:Figure S5}
\end{figure*}

Figure \ref{fig:Figure S5}(a) illustrates the electronic band structure and PDOS for the 1L MoS$_2$/4L PdSe$_2$ HS. In contrast to the MoSe$_2$ case, the valence states of 1L MoS$_2$ (indicated by the red arrow) lie approximately 1 eV below the frontier states of the multilayer PdSe$_2$. This large energy offset facilitates efficient hole transfer from the MoS$_2$ to the PdSe$_2$ layer, leading to the strong PL quenching observed experimentally in the MoS$_2$/PdSe$_2$ control samples (Figure \ref{fig:Figure S5}(b,c) and Figure \ref{fig:Figure S8}).

\clearpage
\widetext

\subsection{\label{PL_633nm} PL spectra under 633 nm excitation}

\begin{figure*}[htbp!]
\centerline{\includegraphics[width=180mm]{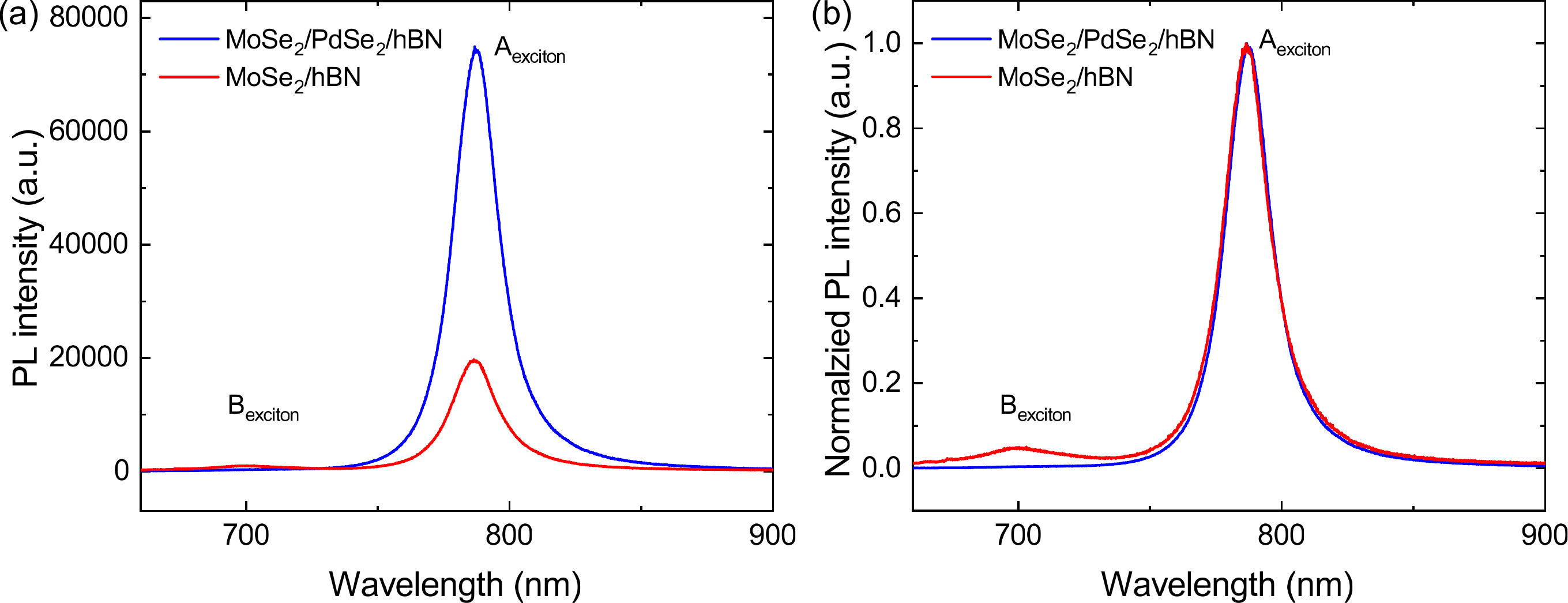}}
\caption{PL spectra of the 1L region and the HS region at 633 nm excitation (a) A-exciton PL enhancement in the HS region (b) B-exciton quenching in the HS region}
\label{fig:Figure S6}
\end{figure*}

Figure \ref{fig:Figure S6} presents PL spectra of the 1L region and the HS region measured under 633 nm excitation. This measurement was performed to verify that the PL enhancement observed under 514 nm excitation is not specific to a single excitation wavelength. A clear enhancement of the A-exciton emission is again observed in the HS region, with the peak intensity increasing by approximately a factor of four relative to bare MoSe$_2$. The persistence of the enhancement under longer-wavelength excitation confirms that the effect is intrinsic to the HS region rather than arising from a particular excitation condition.

\raggedbottom
\newpage

\subsection{\label{SI_temperature}Temperature-dependent evolution of excitonic emission}

\begin{figure*}[htbp!]
\centerline{\includegraphics[width=180mm]{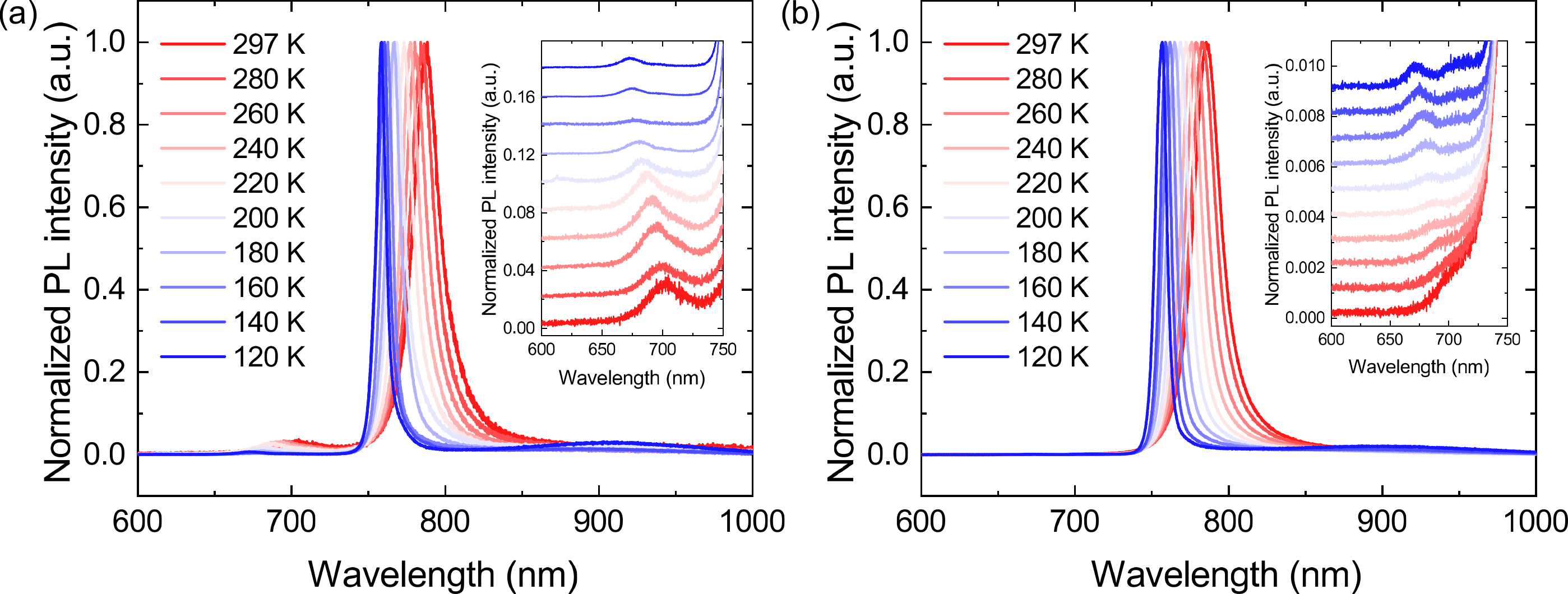}}
\caption{Temperature-dependent normalized PL spectra of (a) the 1L region and (b) the HS region}
\label{fig:Figure S7}
\end{figure*}

Normalized PL spectra of the 1L region (Figure \ref{fig:Figure S7}(a)) were measured from 297 K down to 120 K. With decreasing temperature, the A-exciton peak shows a progressive blueshift of approximately $\Delta$E $\approx$ 61 meV and a reduction in full-width at half-maximum (FWHM) from $\sim$41 meV at 297 K to $\sim$16 meV at 120 K, consistent with the expected temperature dependence of the bandgap and reduced phonon-assisted broadening~\cite{rudin_temperature-dependent_1990}. The relative intensity of the B-exciton, quantified as the ratio I$_B$/I$_A$, decreases from $\sim$0.035 at 297 K to $\sim$0.008 at 120 K, likely due to enhanced B → A exciton intralayer relaxation and the stronger temperature dependence of the A‑exciton radiative efficiency compared to the B‑exciton~\cite{mccreary_-_2018}.

For the HS region (Figure \ref{fig:Figure S7}(b)), the A‑exciton exhibits a blueshift and FWHM narrowing ($\Delta$E $\approx$ 60 meV, from $\sim$42 meV at 297 K to $\sim$19 meV at 120 K), similar to the 1L region. In contrast, the B‑exciton, which is absent at room temperature, gradually emerges at low temperatures (shown in the inset), reaching I$_B$/I$_A$ $\sim$0.001 at 120 K. This behavior reflects the interplay of competing channels: at room temperature, strong interlayer hybridization effectively quenches the B‑exciton and enhances A‑exciton emission, while partially blocking intralayer B → A relaxation. At low temperature, the hybridization weakens due to reduced phonon-mediated coupling, allowing a small fraction of B-excitons to survive and radiatively recombine. Even though B → A relaxation is intrinsically faster at low T, the residual hybridization and suppression of EEA at low powers enable the appearance of a weak B-exciton emission. Notably, the B-exciton in the HS region remains much weaker than in the 1L region, highlighting the persistent influence of interlayer hybridization on exciton dynamics.

\raggedbottom
\newpage

\subsection{\label{PLE_absorbance}PLE spectra MoS$_2$-based samples}

\begin{figure*}[htbp!]
\centerline{\includegraphics[width=80mm]{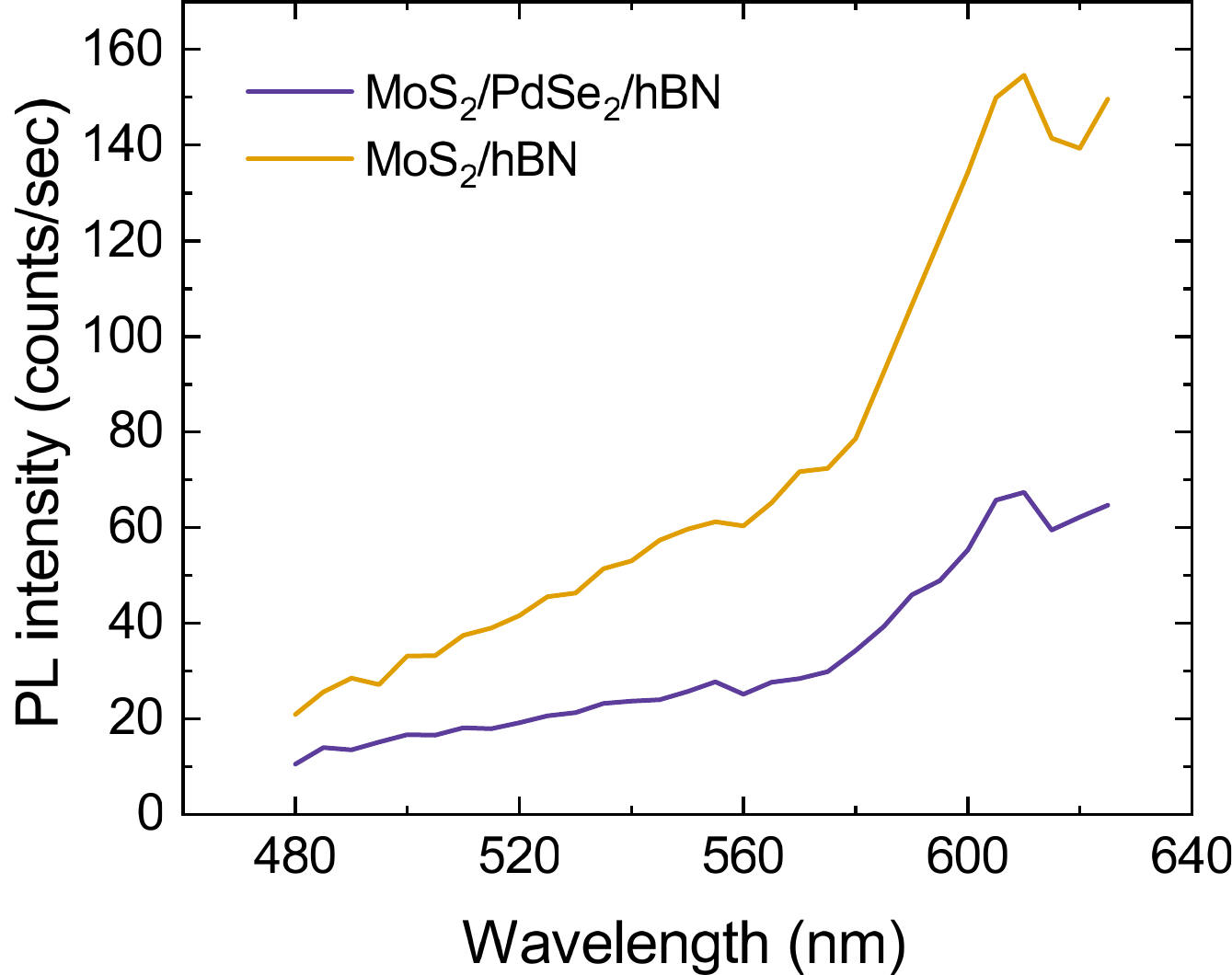}}
\caption{PLE spectra of the 1L region and the HS region}
\label{fig:Figure S8}
\end{figure*}

Figure \ref{fig:Figure S8} presents the PLE spectra for the MoS$_2$-based control samples. In contrast to the broadband enhancement observed in the MoSe$_2$ system, the HS region (purple line) exhibits a systematically reduced PL intensity compared to the isolated 1L MoS$_2$ (yellow line) across the entire excitation range (480–625 nm), with the emission intensity lower by a factor of $\sim$1.8–2.5 depending on excitation wavelength.

\clearpage
\widetext

\end{document}